\documentclass[prl,aps,showpacs,twocolumn,draft]{revtex4}
\usepackage{epsf}
\begin{document}
\title{Exact microscopic wave function for a topological quantum membrane}
\author{Shou-Cheng Zhang}

\address{Department of Physics, McCullough Building,
Stanford University,
Stanford,  CA~~94305-4045}

\begin{abstract}
\par~\par
The higher dimensional quantum Hall liquid constructed recently
supports stable topological membrane excitations. Here we
introduce a microscopic interacting Hamiltonian and present its
exact ground state wave function. We show that this microscopic
ground state wave function describes a topological quantum
membrane. We also construct variational wave functions for
excited states using the non-commutative algebra on the four sphere.
Our approach introduces a non-perturbative method to quantize
topological membranes.
\end{abstract}
\pacs{73.43.-f,11.25.-w,87.16.Dg} \maketitle

Recently, a higher dimensional generalization of the quantum Hall
effect has been constructed by Zhang and Hu (ZH)\cite{zhang2001A}. The
fundamental fermionic fluid particles move on the surface of a
four sphere ($S^4$)with radius $R$, and carry $SU(2)$ gauge
degrees of freedom in the representation $I$. The instanton
density of the $SU(2)$ gauge field is uniformly distributed over
$S^4$. ZH considered the limit where $I\rightarrow \infty$ when
$R\rightarrow \infty$, such that $R^2/2I\equiv l^2$ is held
constant. $l$ defines the fundamental magnetic length in this
problem. This quantum Hall liquid shares many properties of the
familiar 2D quantum Hall liquid, including incompressibility,
fractional charge and gapless edge states. This theory has been
further developed in
refs. \cite{hu2002A,fabinger2002A,karabali2002A,chen2002A,sparling2002A,kimura2002A,bernevig2002A,ramgoolam2002A,elvang2002A,chen2002B}.

Bernevig et al\cite{bernevig2002A} have recently constructed a
topological field theory for the new quantum Hall liquid. The
configuration space in this problem is $CP_3$, which is locally
the product of the orbital space $S^4$ and the isospin space
$S^2$. The Chern-Simons theory can be defined either over the
configuration space $CP_3$, or on the orbital space $S^4$. The
later can be obtained from the former through the fuzzification of
the isospin sphere $S^2$. This field theoretical study reveals an
important class of extended topological objects, including the
membrane (2-brane) and the 4-brane. The membranes wrap the
isospin $S^2$ and have a non-trivial statistical interaction which
generalize the concept of fractional statistics of Laughlin
quasi-particles.

In this paper, we investigate microscopic properties of the
membranes found in the study of Bernevig et al. We shall introduce
a microscopic interaction Hamiltonian in the lowest-Landau-level (LLL),
and find its {\it exact} ground state wave function. This wave
function is a natural generalization of Laughlin's wave function
for the 2D QHE\cite{laughlin1983A,haldane1983A}. We then show
that this wave function describes a collection of particles forming
a membrane, wrapped around the isospin $S^2$. This wave function
is a $SO(5)$ singlet, therefore, the center of mass of the membrane
is uniformly delocalized on $S^4$. We also discuss the excitations
of the membrane in terms of the non-commutative algebra on $S^4$.

The importance of the exact membrane wave function may be viewed from
different perspectives. Up to this point, only microscopic information
about the ZH model are based on non-interacting physics. An elementary
analysis of the boundary degrees of freedom reveals an ``embarrassment
of riches"\cite{zhang2001A,hu2002A}. The total configuration space at the boundary is
$S^3\times S^2$. In the non-interacting limit, isospin excitations are
gapless, leading to massless states with all helicities. In addition,
there is also an incoherent fermionic background. The entropy at the
boundary therefore scales like $R^5$, unlike the $R^3$ scaling one
would expect from a conventional $3+1$ dimensional field theory. A
possible solution to this problem could come from an ``isospin gap",
introduced by the mutual interaction among the particles. In this case,
for energy scales below the ``isospin gap", the entropy
would scale like $R^3$. Our exact membrane solution in the presence
of the interaction indeed suggests this behavior. Since our membrane
wraps the isospin $S^2$, its internal degrees of freedom behaves like a
a 2D QH liquid with an incompressibility gap. Beyond the problem of
current interest, our exact membrane solution gives a new way to
quantize a membrane beyond Polyakov's path integral quantization,
and could yield valuable information about the strong coupling limit of
quantum membranes. Finally, two possible interpretation of the
elusive ``M theory" are the matrix and membrane
theories\cite{banks1997A,taylor2001A}. In our
model, these two theories are intimately related. In the LLL, the
fundamental particles can be described by a matrix model. Our exact
wave function shows microscopically how membranes emerge from a collection
of matrix particles.

Let us first recall that the single particle wave function in the LLL
of the 4DQHE problem is given by
\begin{eqnarray}
\sqrt{\frac{p!}{m_1! m_2! m_3! m_4!}}
\Psi_1^{m_1} \Psi_2^{m_2} \Psi_3^{m_3} \Psi_4^{m_4}
\label{LLL}
\end{eqnarray}
with integers $m_1+m_2+m_3+m_4=p=2I$. The four component spinors
$\Psi_\alpha$ can be expressed directly in terms of the $S^4$ and
the $S^2$ coordinates, as given in eq. (5) of ref. \cite{zhang2001A}.
The degeneracy of the single particle ground states is given by
\begin{eqnarray}
D(p)=\frac{1}{6}(p+1)(p+2)(p+3)
\label{degeneracy}
\end{eqnarray}
A rather remarkable feature is that while higher LL states are
$SO(5)$ symmetric, the LLL states enjoy an additional $SU(4)$
symmetry, as one can see directly from eq. (\ref{LLL}).
$SO(5)$ group is isomorphic to the $Sp(4)$ group, which differs
from the $SU(4)$ group only through an additional structure associated
with the charge conjugation matrix ${\cal R}$. Let $\Gamma_a$, with
$a=1,..,5$ be the five Dirac Gamma matrices satisfying the
Clifford algebra $\{\Gamma_a, \Gamma_b\}=2\delta_{ab}$, then
$\Gamma_{ab}=-\frac{i}{2} [\Gamma_a, \Gamma_b]$ form the generators
of the $SO(5)$ Lie algebra. The ${\cal R}$ matrix is defined by the
following properties:
\begin{eqnarray}
  {\cal R}^2 = -1, \ \ \ {\cal R}^{\dagger} =  {\cal R}^{-1} = {}^t {\cal R} = -{\cal R}   \\
  {\cal R}\,\Gamma^a {\cal R} = -{}^t\Gamma^a, \ \ \ {\cal R}\,\Gamma^{ab}{\cal R} = {}^t\Gamma^{ab}
\label{R_prop}
\end{eqnarray}
The relation
$ {\cal R}\,\Gamma^{ab}{\cal R}^{-1} = -(\Gamma^{ab})^*$ indicates that the spinor
representation of $SO(5)$ is pseudo-real. The ${\cal R}$ matrix plays a role similar
to that of $\epsilon_{\alpha\beta}$ in $SO(3)$. In the explicit representation
given by eq. (4) of ref. \cite{zhang2001A}, the ${\cal R}$ matrix takes the form
\begin{eqnarray}
{\cal R} = -i \left( \begin{array}{cc}
                     \sigma_y  &  0  \\
                    0  &  \sigma_y  \end{array} \right)
\label{R_explicit}
\end{eqnarray}
where $\sigma_y$ is a Pauli matrix.
The presence of the ${\cal R}$ matrix breaks the $SU(4)$ symmetry down to
the $SO(5)=Sp(4)$ symmetry. Since the LLL wave functions do not
involve the ${\cal R}$ matrix, they are $SU(4)$ invariant. However, the
$R$ matrix is needed to construct wave functions in higher LLs\cite{chern2002A},
which are only $SO(5)$ invariant.

Having reviewed the wave functions in the LLL and introduced the concept
of the ${\cal R}$ matrix, we now present our microscopic wave function.
In the 2DQHE, the $\nu=1$ wave function can be expressed either as a
Slater determinant or as a Jastrow-Laughlin type product wave function. The
van der Monde identity relates them exactly. However, this identity
does not hold in higher dimension. We shall see the profound physical
implications introduced by this inequivalence. In ref. \cite{zhang2001A},
ZH constructed many-body wave functions by using the
Slater determinant, and following Laughlin, by taking odd powers of these
Slater determinant. They showed that these wave functions describe
incompressible quantum liquids. However, one could proceed in a different
way here, by constructing wave functions using the inequivalent
Jastrow-Laughlin product form. Such a wave function takes the
form
\begin{eqnarray}
\Phi_0=\prod_{i<j} (\Psi_\alpha (i) {\cal R}^{\alpha\beta}
\Psi_\beta (j))^m
\label{Psi}
\end{eqnarray}
where $m$ is an odd integer and $i,j$ refers to $i$th and $j$th particles
in the system. If we replace ${\cal R}^{\alpha\beta}$
by $\epsilon^{\alpha\beta}$, and let $\alpha,\beta$ to take values
of $1,2$, this would transform exactly into Laughlin's
wave function expressed in Haldane's sphereical geometry\cite{haldane1983A}.
Our wave function has the following properties:

1) When $m$ is an odd integer, this wave function is antisymmetric when
particle coordinates are exchanged. Therefore, this
wave function describes a fermionic system.

2) The wave function is a $SO(5)$ singlet. This is because
every term in the product,
$\Psi_\alpha (i) {\cal R}^{\alpha\beta}\Psi_\beta (j)$
is a $SO(5)$ singlet, by virtue of eq. (\ref{R_prop}).

Since the wave function involves the ${\cal R}$ matrix
explicitly, the symmetry in the LLL is broken from
$SU(4)$ down to $SO(5)$.

3) When the product is expanded, the spinor coordinate $\Psi(i)$
occurs $m(N-1)$ times. Therefore, the wave function for the
$i$th particle takes the form of (\ref{LLL}), with
$p=m(N-1)$. ZH showed that single particle level spacing
becomes finite in the limit when $p/R^2=1/l^2$ is held
constant. Therefore, the number of particles in the wave function
(\ref{Psi}) scales like $N\sim p\sim R^2$. In other words,
{\it the wave function (\ref{Psi}) describes a two dimensional
object}.

This wave function can be represented graphically. We associate
each particle with $p$ dots, representing a symmetric spinor
state of the form of (\ref{LLL}). We draw a solid line representing
a {\it contraction} between the $i$th and the $j$th particle
through the ${\cal R}$ matrix. The resulting graphical representation
for $N=4$, $m=1,3$ is depicted in Fig. 1a and Fig. 1b.

\begin{figure}[h]
\centerline{\epsfysize=4cm \epsfbox{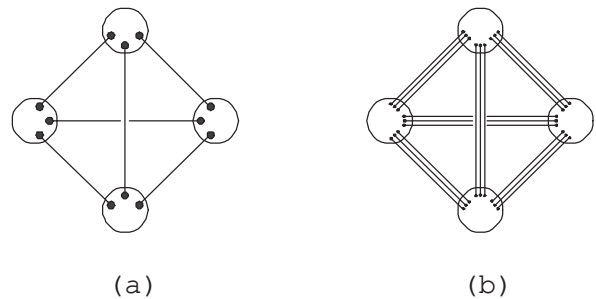}} \caption{ A
graphical representation of the wave function (\ref{Psi}), when
the product is expanded. Each particle is
denoted by a circle with $p$ dots, representing a LLL state
$(p,0)$. Each solid line denotes a $SO(5)$ singlet bond formed
between the spinor indices of particles $i$ and $j$.}
 \label{fig1}
\end{figure}

4) In order to see what kind of two dimensional object is
described by the correlated wave function (\ref{Psi}), we borrow
from Laughlin's plasma analogy. $\int \prod_i dX_i dn_i |\Phi_0|^2$
can be interpreted as the partition function of a classical gas,
living in the six dimensional $CP_3$ configuration space. The
Boltzmann weight for this classical gas is
\begin{eqnarray}
|\Phi_0|^2 = \prod_{i<j} |\Phi(ij)|^{2m}=e^{m \sum_{i<j}
Log|\Phi(ij)|^2}
\label{gas}
\end{eqnarray}
where $\Phi(ij)=\Psi_\alpha (i) {\cal R}^{\alpha\beta} \Psi_\beta
(j)$. Since the wave function involves only pair-wise
correlations, we see that the classical gas interacts via a
two-body potential only. $m$ can be interpreted as the effective
inverse temperature, $\beta=1/kT$. Using the tensor identity
\begin{eqnarray}
{\cal R}_{\alpha\beta} {\cal R}_{\gamma\delta} = \frac{1}{4}
\delta_{\alpha\gamma} \delta_{\beta\delta} + \frac{1}{4}
\Gamma^a_{\alpha\gamma} \Gamma^a_{\beta\delta} - \frac{1}{4}
\Gamma^{ab}_{\alpha\gamma} \Gamma^{ab}_{\beta\delta}
\label{tensor}
\end{eqnarray}
we obtain the explicit form of the pair-wise potential
\begin{eqnarray}
|\Phi(ij)|^2= \frac{1}{4} + \frac{1}{4} X_a(i) X_a(j) -
\frac{1}{4} F_{ab}^\alpha(i)n_\alpha(i) F_{ab}^\beta(j)n_\beta(j)
\label{Phi2}
\end{eqnarray}
here $X_a(i)$, with $X_a^2=1$, describes the coordinate of the
$i$th particle on the orbital space $S^4$, and $n_\alpha(i)$, with
$n_\alpha^2=1$, describes its coordinate on the isospin  space
$S^2$. $F_{ab}^\alpha$ is the Yang-Mills field strength of the
$SU(2)$ instanton over $S^4$, explicitly given by eq. (19) of
ref. \cite{bernevig2002A}.

From the sign of the second terms in eq. (\ref{Phi2}), we see that
the gas particles interact via an {\it attractive} interaction on
$S^4$. This is in direct contrast to Laughlin's plasma, where the
gas particles interact via a {\it repulsive} interaction on
$S^2$. Therefore, at low temperature, our gas particles have a
natural tendency to cluster to the same point on $S^4$. However,
over every point on $S^4$, there is also a large, internal
isospin $S^2$ for the gas particles to ``live". From the sign of
the third term in eq. (\ref{Phi2}), we see that the gas particles
{\it repel} each other on the isospin $S^2$, just like the case
of the Laughlin plasma. From these observations, we see that our
gas particles cluster to the same point on $S^4$, but {\it
uniformly} fill the isospin $S^2$ like a 2D QH liquid. Viewed
from the whole $CP_3$ point of view, the gas particles form a 2D
{\it membrane}, wrapped around the isospin sphere $S^2$. The
center-of-mass of the membrane is a point on $S^4$, our wave
function consists of an equal weight linear superposition of the
center of mass position over $S^4$, therefore, the ground state
is a $SO(5)$ singlet.

Having exhibited the key properties of our wave function and its
classical plasma analog, we now present a microscopic quantum
interaction Hamiltonian for which it is an exact ground state.
Our construction follows closely Haldane's pseudo-potential
formalism\cite{haldane1983A}. Since every particle is in the $(p,0)$ irreducible
representation (irreps) of $SO(5)$, the total $SO(5)$ quantum
number of a pair of particles $i$ and $j$ is {\it generically}
given by
\begin{eqnarray}
(p,0)\bigotimes (p,0)=\sum_{k=0}^p \sum_{l=0}^k (k+l,k-l)
\label{total}
\end{eqnarray}
On the other hand, in our wave function (\ref{Psi}), there are
already $m$ $SO(5)$ singlet contractions between the particle $i$
and $j$. Therefore, the total $SO(5)$ quantum number between
these two particles {\it in our wave function} can only be
contained in $(p-m,0)\bigotimes(p-m,0)$. Based on this observation, we
introduce the projector Hamiltonian in the LLL as $H=\sum_{i<j} H_{ij}$,
where
\begin{eqnarray}
H_{ij}=\sum_{k=p-m+1}^p
\sum_{l=0}^k \lambda(k+l,k-l) P_{ij}(k+l,k-l)
\label{H}
\end{eqnarray}
here $P_{ij}(k+l,k-l)$ is a projection operator onto the $SO(5)$
irreps $(k+l,k-l)$ between a pair of particles $i$ and $j$ and
$\lambda(k+l,k-l)>0$ is the interaction parameter in a given
channel. This Hamiltonian operates fully within the LLL and is
positive definite. Since a pair of particles in our wave function can not
have any of the $SO(5)$ irreps specified in (\ref{H}), it is
annihilated by all projectors. Therefore, we have proven that our
wave function (\ref{Psi}) is an eigenstate of the Hamiltonian
(\ref{H}) with zero eigenvalue. Since the Hamiltonian is also
positive definite, our wave function must therefore be a ground
state. Our experience leads us to conjecture that this is also an
unique ground state in the spherical geometry.

Having shown the exact ground state wave function (\ref{Psi})
of the interacting Hamiltonian (\ref{H}), we now proceed to discuss
the excited states. Following Feynman's construction of elementary
excitations with liquid helium\cite{feynman1954A}, we try to construct variational
states of the form:
\begin{eqnarray}
\Phi=\prod_{i} F(X_i,n_i) \Phi_0
\label{feynman}
\end{eqnarray}
where $F(X)$ is a single particle wave function over $S^4$. In Feynman's
case, he simply took $F(x)$ to be plane waves, and his wave
function describes the center-of-mass motion of a correlated quantum
liquid with finite momentum. In our case, one could use the
spherical harmonics over $S^4$ for $F(X)$:
\begin{eqnarray}
F(X)=\sum_{L={l_1,..,l_5}} f_L X^L, \ X^L\equiv X_1^{l_1} X_2^{l_2}
X_3^{l_3} X_4^{l_4} X_5^{l_5}
\label{F}
\end{eqnarray}
Here $l_1+..+l_5=l$, and $f_L$ is choosen such that $F$ belongs
to the fully symmetric traceless tensor representation $(l,l)$
of the $SO(5)$ group. We argued earlier that the membrane wave function
$\Phi_0$ is a $SO(5)$ singlet, which means that the center-of-mass of the
membrane has the lowest $SO(5)$ angular momentum on $S^4$. The
more general wave function $\Phi$ given in (\ref{feynman}) describes
higher angular momentum of the center-of-mass on $S^4$.

However, there is a serious problem with the function $F(X)$.
Since $X_a=\bar\Psi \Gamma_a\Psi$, $F(X)$ depends both on
$\bar\Psi$ and $\Psi$. But eq. (\ref{LLL}) shows that the single
particle wave functions in the LLL can only involve $\Psi$ but not
$\bar\Psi$. The solution of this problem is provided by Girvin,
MacDonald and Platzman\cite{girvin1986A}. One simply
needs to use the projection of $X_a$ in the LLL, which is
\begin{eqnarray}
X_a=\frac{1}{p}\Psi \Gamma_a \frac{\partial}{\partial\Psi}
\label{X}
\end{eqnarray}
The effect of $X_a$ operating on a $SO(5)$ singlet bond formed
by ${\cal R}$ is to turn it into a vector bond ${\cal R}\Gamma_a$.
After the projection, $X_a$'s become operators
and no longer commute with each other, in fact, they satisfy the
non-commutative algebra outlined in \cite{zhang2001A}. As a
consequence of the non-commutativity, $F(X)$ does not only
give the fully symmetric traceless tensor representation $(l,l)$
of the $SO(5)$ group, but includes all $SO(5)$ irreps in the series
$\underline{\bf 5}\bigotimes\underline{\bf 5}\bigotimes\underline{\bf 5}...$
A physical interpretation of this result is that
because of the non-commutative geometry in the LLL, the center-of-mass
degrees of freedom are coupled to the internal membrane degrees
of freedom. A full calculation of the
variational energies for the wave function (\ref{feynman}), with
$F$ given by (\ref{F}) and (\ref{X}) will be carried out in the
future, possibly with the assistance of numerical calculations.

However, even without explicit calculations, we can anticipate the
result based on our experience with the 2D QHE. We argued before
that the our membrane wrapping the isospin $S^2$ is made out of
a 2D QH liquid, which has an incompressibility gap. Therefore, it
appears likely that our quantum membrane does not have the ``spike
instability" of a classical membrane\cite{taylor2001A}, where
arbitrarily long spikes can be created at low energies.
This picture has important implications on the relativistic edge
dynamics of the 4DQH liquid. In the non-interacting problem, the
entropy at the edge scale like $R^3\times R^2=R^5$, since the
internal isospin excitations are gapless. In this work, we have
seen that interaction can introduce a high degree of correlation.
In the strong coupling limit, there are no free particle excitations,
only correlated membrane excitations. Furthermore, the membranes
wrap the isospin $S^2$ by forming a 2D QH liquid, which has an
incompressibility gap. In this case, for energies below the
incompressibility gap, the effective entropy at the edge would
scale like $R^3$. {\it This effect gives a mechanism of ``dynamical
dimensional reduction".}
Different internal membrane excitations appear
as different helicity states in the $3+1$ dimensional world view.
Therefore, higher helicity states would naturally acquire an energy gap, as a
result of the interaction and quantum correlations built into the
membrane wave function. However, we do not yet know a natural
mechanism within our framework to gap only states with helicities
greater than three. Nonetheless, we believe that the exact membrane
wave function represented in this paper provides a key step towards
understanding the strong correlation effects in the 4D QHE model.

We would
like to acknowledge useful discussions with A. Bernevig, C. H.
Chern, J.P. Hu and N. Toumbas. This work is supported by the NSF
under grant numbers DMR-9814289.

\bibliographystyle{prsty}
\bibliography{brane}

\end{document}